\documentclass[aps,pre,twocolumn,superscriptaddress,10pt]{revtex4-1}
\usepackage{graphicx}
\usepackage{bm}
\usepackage[usenames]{color}
\bibstyle{apsrev.bib}

\usepackage{epsfig}
\usepackage{amsmath}
\DeclareMathOperator{\Tr}{Tr}
\usepackage{amssymb}
\usepackage{array}
 
\newcommand{\be}{\begin{equation}}
\newcommand{\ee}{\end{equation}}
\newcommand{\beqn}{\begin{eqnarray}}
\newcommand{\eeqn}{\end{eqnarray}}

\usepackage{dsfont}

\begin{document}

\title{Testing the validity of random-singlet state for long-range hopping models through the scaling of entanglement entropy}

\author{R\'obert Juh\'asz}
\email{juhasz.robert@wigner.hu}
\affiliation{Wigner Research Centre for Physics, Institute for Solid State Physics and Optics, H-1525 Budapest, P.O. Box 49, Hungary}

\date{\today}

\begin{abstract}
We consider a sublattice-symmetric free-fermion model on a one-dimensional lattice with random hopping amplitudes decaying with the distance as $|t_l|\sim l^{-\alpha}$, and address the question how far an analogue of the random-singlet state (RSS) conceived originally for describing the ground state of certain random spin chains is valid for this model. For this purpose, we study the effective central charge characterizing the logarithmic divergence of the entanglement entropy (EE) and the prefactor of the distribution of distances between localization centers on the two sublattices, which must fulfill a consistency relation for a RSS. For $\alpha>1$, we find by exact diagonalization an overall logarithmic divergence of the entanglement entropy with an effective central charge varying with $\alpha$. The large $\alpha$ limit of the effective central charge is found to be different from that of the nearest-neighbor hopping model. 
The consistency relation of RSS is violated for $\alpha\le 2$, while for $\alpha>2$ it is possibly valid, but this conclusion is hampered by a crossover induced by the short-range fixed point.
The EE is also calculated by the strong-disorder renormalization group (SDRG) method numerically. Besides the traditional scheme, we construct and apply a more efficient minimal SDRG scheme having a linear (nearest-neighbor) structure, which turns out to be an accurate approximation of the full SDRG scheme for not too small $\alpha$. The SDRG method is found to provide systematically lower effective central charges than exact diagonalization does, nevertheless it becomes more and more accurate for increasing $\alpha$. Furthermore, as opposed to nearest-neighbor models, it indicates a weak dependence on the disorder distribution.        
\end{abstract}

\maketitle


\section{Introduction}

The localization of states in disordered systems has attracted much interest since the pioneering work of Anderson \cite{anderson}. Among the first works in this field, it was pointed out that a random potential causes a localization of all eigenstates in the one-dimensional tight-binding model independently of the strength of disorder \cite{mt}. Since then the study of localization was extended to various directions such as higher dimensions, long-range hopping, random hopping amplitudes, or interacting models, just to mention a few \cite{50,tarquini,economou,mbl}. 
In this paper we focus on the effect of long-range hopping with an amplitude decaying with the distance algebraically as $|t_l|\sim l^{-\alpha}$ where $\alpha$ is regarded as a tunable parameter. Localization in such systems, which occurs for $\alpha>1$ \cite{mirlin,lima}, can be defined in a weaker sense than in short-range systems: instead of an exponential localization, the wave function has always an algebraic envelope $|\psi_l|\sim l^{-\alpha}$ here.      
From a more general perspective, disordered quantum systems with long-range interactions appear in several physical contexts, for instance in doped semiconductors in which magnetic impurities interact via long-range exchange couplings \cite{rk,kasuya,yosida,mott}. 
  
An important cornerstone of the theory of random quantum systems is the strong-disorder renormalization group (SDRG) method, which was originally formulated and solved for random antiferromagnetic (AF) spin chains \cite{mdh,fisherxx,im}. 
The ground state obtained by this approach is a peculiar product state comprising pairs of spins in a singlet state, where the constituents of singlets can be arbitrarily far away from each other, with a distribution of distances $l$ decaying as $p_l\simeq \frac{2}{3}l^{-2}$ \cite{refael,hoyos2007,juhasz2021}. 
Although the random-singlet state is an approximation, the perturbative SDRG scheme being asymptotically exact at low energies, it still captures the low-energy (large-scale) properties correctly, such as the asymptotics of spatial correlation functions or the entanglement entropy of contiguous spin blocks \cite{fisherxx,refael,laflorencie,yang}.     
The XX spin chain with arbitrary couplings is well-known to be closely related to free (non-interacting) fermions via the Jordan-Wigner transformation. 
In accordance with this exact relation, an SDRG scheme can also be formulated directly for a system of lattice fermions hopping with random transition amplitudes \cite{motrunich,melin,hoyos,junction}, and this puts localization problems in fermion systems in an interesting perspective.  
For bipartite lattices and in the absence of on-site potentials (i.e. for a pure off-diagonal disorder), the approximate one-particle states of fermion models obtained by the SDRG method are states which are perfectly localized on two sites $i$ and $j$ belonging to different sublattices: $\psi^{\rm SDRG}_k(n)=\frac{1}{\sqrt{2}}(\delta_{in}-\delta_{jn})$. These states are analogous to the singlet pairs in the random-singlet state of AF spin chains, therefore we will adopt the terminology of AF spin chains in this paper and call the pair-localized (half-filled) ground state of the hopping model obtained by the SDRG method as a random-singlet state (RSS). 
Using the distribution of ``singlet lengths'', $l=|i-j|$, and the relation between $l$ and the energy $E$ of eigenstates, $|\ln|E||\sim \sqrt{l}$ provided by the SDRG method \cite{fisherxx}, one readily obtains the form of the Dyson singularity in the density of states around the band center ($E=0$): $\rho(E)\sim |E|^{-1}|\ln|E||^{-3}$ \cite{dyson,eggarter}. 
Although the true eigenstates are not perfectly localized on two sites but have a finite extension around them (for non-zero energies),  the distance $l=|i-j|$ between localization centers on odd and even sublattices
has recently been confirmed to follow the asymptotic distribution $p_l\simeq \frac{2}{3}l^{-2}$ predicted by the SDRG method \cite{juhasz2021}. 

In this paper, we address the question how stable the random-singlet state is against long-range hopping and whether the approximate state obtained by the SDRG method is appropriate to describe large-scale properties such as the asymptotics of the entanglement entropy. 
This or related questions have already been studied in various models. In a random-bond hopping model with sublattice symmetry the correlations were studied and $\alpha=2$ was found to be a critical value above which the decay exponent of the correlation function remained unchanged from the nearest-neighbor model \cite{bhatt}. Based on the numerical results, the random singlet phase in long-range spin chains was conjectured to be stable for $\alpha>2$. In Ref. \cite{roy}, the entanglement entropy was studied in XX and free-fermion chains with random long-range couplings and, as opposed to the RSS of short-range models in which the entanglement entropy diverges logarithmically with the block size as $S_{\ell}\simeq\frac{\ln 2}{3}\ln\ell$, it was found to increase as a power of $\ell$ for $\alpha<1$. Similar power-law violation of the entanglement area law was found for $\alpha<1$ in the power-law random banded model which contains both diagonal and off-diagonal disorder \cite{yang_2014}.   
In bond-disordered XX chains with long-range couplings, the entanglement entropy was studied by the SDRG method, by density-matrix renormalization group and by exact diagonalization in small systems \cite{mohdeb}. The divergence was found to be logarithmic by all methods; the SDRG method provided the same prefactor $\frac{\ln 2}{3}$ as in the short-range model, independently of $\alpha$, while exact diagonalization data showed a larger prefactor with a weak $\alpha$-dependence, but the asymptotic region was presumably not reached here.     

We shall consider in this paper a one-dimensional hopping model with pure off-diagonal disorder by adding long-range hopping with real, random amplitudes decreasing algebraically with the distance on average as $|t_l|\sim l^{-\alpha}$. We allow hopping only between odd and even sublattices to keep the sublattice (or chiral) symmetry of the nearest-neighbor model, as was done in Ref. \cite{bhatt}. 
Although the exact relationship with the XX spin model is lost owing to the presence of long-range terms, the system of non-interacting fermions under study can still be regarded as a simple model of AF spin systems, and we expect it to have qualitatively similar properties to the latter.
The advantage of the hopping model against AF spin models is that relatively large systems can be numerically studied by exact diagonalization, allowing for a more precise comparison with the SDRG approximation. 
We study the block size dependence of the entanglement entropy by exact diagonalization and by the SDRG method numerically. In addition to this, we also calculate the distribution of the distances between localization centers on odd an even sublattices and, checking the consistency with entanglement entropy divergence, we investigate the question whether the long-range model has a generalized RSS with a singlet-length distribution possibly different from that of the short-range model. 

Our main results are summarized as follows.     
The entanglement entropy increases logarithmically with the subsystem size in the extensive regime $\alpha>1$. The prefactor in front of the logarithm differs from that of the short-range model, $\frac{\ln 2}{3}$ and varies with $\alpha$. 
Moreover, the SDRG method also shows a weak dependence on the disorder distribution.
For $\alpha\le 2$ the singlet-length distribution and entanglement entropy scaling are incompatible with a RSS, while for $\alpha>2$, they are possibly consistent and compatible with a generalized, non-universal RSS. This conclusion is, however, hampered by the disturbance of the short-range fixed point.
The SDRG method, although relatively accurate for not too small $\alpha$, does not reproduce correctly the non-universal prefactor of the entanglement entropy. 
         
The paper is organized as follows. 
In Sec. \ref{sec:model} the model is defined and its solution is formulated as a singular-value decomposition. The traditional SDRG method is introduced and an efficient minimal SDRG scheme is developed in Sec. \ref{sec:sdrg}. 
 Sec. \ref{sec:conjectures} presents simple reasonings based on the SDRG approach, which lead to the non-universality with respect to the disorder distribution and to the presence of a crossover effect caused by the short-range fixed point.  
Numerical results are presented in Sec. \ref{sec:numerical} and discussed in \ref{sec:discussion}. 

\section{The model}
\label{sec:model}
 
We consider a system of non-interacting fermions hopping on a one dimensional lattice with an even number $L$ of sites and with periodic boundary condition. The Hamiltonian of the model is 
\be
H=\sum_{i,j=1}^{L/2}t_{ij}(c_{2i-1}^{\dagger}c_{2j}+c_{2j}^{\dagger}c_{2i-1}),
\label{hamilton}
\ee
where $c_n^{\dagger}$ and $c_n$ denote fermion creating and annihilating operators on site $n$, respectively. As can be seen there are no on-site terms in the Hamiltonian and hopping is allowed only between sites belonging to different (odd and even) sublattices. As a consequence, the model has a chiral symmetry, which manifests itself in that the one-particle states appear in pairs with energies $\pm E$ and the corresponding eigenstates differ only by the sign of wave function components on one of the sublattices. 
The hopping amplitudes are chosen to be of the form 
\be
t_{ij}=w_{ij}l_{ij}^{-\alpha}, 
\ee
where 
$l_{ij}=\min\{|n-m|,L-|n-m|\}$ is the shortest distance between sites $n=2i-1$ and $m=2j$ on the periodic lattice and $\alpha>0$ is a tunable parameter controlling the range of the long-range hopping. In the present paper, we shall restrict ourselves to the extensive regime, $\alpha>1$. The factors $w_{ij}$ are real, independent and identically distributed random variables.

\subsection{Solution by singular-value decomposition} 

We consider ground-state properties of the model at half filling, where all one-particle states with a negative energy are occupied. The bipartite structure of the model can be used to reduce the computational demand of determining the one-particle eigenstates. 
Introducing the matrix $T$ with elements $T_{ij}=t_{ij}$, $i,j=1,2,\dots,L/2$, and the normalized column vectors $\phi_k=\sqrt{2}[v_k(1),v_k(3),\dots,v_k(2i-1),\dots]^T$ and $\psi_k=\sqrt{2}[v_k(2),v_k(4),\dots,v_k(2i),\dots]^T$ of the odd and even components of the $k$th (real) eigenvector $v_k=[v_k(1),v_k(2),\dots]^T$, respectively, the eigenvalue problem of the model is of the form 
\beqn
T\psi_k=E_k\phi_k 
\label{ev1a}
 \\
T^T\phi_k=E_k\psi_k.
\label{ev1b}
\eeqn
These equations can also be rewritten as separate eigenvalue equations 
for $\phi_k$ and $\psi_k$ in the form 
\beqn
T^TT\psi_k=E_k^2\psi_k \nonumber \\
TT^T\phi_k=E_k^2\phi_k.
\label{ev2}
\eeqn
This reduction of the original Hamilton matrix of size $L\times L$ to matrices of order $L/2$ was also applied in Ref. \cite{bhatt}. 
Here we make the observation that this formulation of the eigenvalue problem is equivalent to the singular-value decomposition of matrix $T$. 
Forming $L/2\times L/2$ matrices from the positive-energy column vectors as
$\Phi=(\phi_1,\phi_2,\dots,\phi_{L/2})$ and $\Psi=(\psi_1,\psi_2,\dots,\psi_{L/2})$, Eq. (\ref{ev1a}) assumes the form 
\be
T\Psi=\Phi D,
\label{ev_matrix}
\ee
where $D={\rm diag}\{E_1,E_2,\dots,E_{L/2}\}$ is a diagonal matrix with the positive eigenvalues in the diagonal. 
From Eq. (\ref{ev2}) and the fact that $\phi_k$ and $\psi_k$ are normalized, one can see that $\Phi$ and $\Psi$ are orthogonal: 
$\Phi^T\Phi=\Psi^T\Psi=1$. Using this, Eq. (\ref{ev_matrix}) can be recast as 
\be
T=\Phi D\Psi^T.
\ee
This is nothing but the singular-value decomposition of $T$, where the positive eigenvalues are the singular values and the columns of $\Phi$ and $\Psi$ ($\phi_k$ and $\psi_k$) are the left- and right-singular vectors of $T$, respectively.
The eigenstates with a negative energy are then given by the components $\frac{1}{\sqrt{2}}\phi_k$ and  $-\frac{1}{\sqrt{2}}\psi_k$. 
We note that $\phi_k^T\phi_k=\psi_k^T\psi_k$ holds for each eigenstate, i.e. the particle can be found on the two sublattices with an equal probability.   

\subsection{Localization centers and entanglement entropy}

In the nearest-neighbor hopping model with off-diagonal disorder, the low-energy eigenstates are localized in two spatially separated regions which are supported by different sublattices. The SDRG method provides idealized states which are perfectly localized on a single site of each of these two regions. The positions of these two sites are correlated since, according to the SDRG method, their separation $l$ has an asymptotic probability distribution \cite{refael,hoyos2007,juhasz2021}:
\be
p_l\simeq \frac{2}{3}l^{-2}.
\label{pl}
\ee
Thus larger separations are less probable. Note that the distances $l$ take on odd values only. 
Regarding the true eigenstates, the localization is not perfect, and the best one can do is to consider the localization centers $2i_k-1$ and $2j_k$ on the odd and even sublattices, respectively, which can be defined by the maxima of the eigenvector components:     
\beqn 
|\phi_k(i_k)|&=&\max_n\{|\phi_k(n)|\}  \nonumber \\
|\psi_k(j_k)|&=&\max_n\{|\psi_k(n)|\}.         
\label{maxima}
\eeqn
The distances between localization centers has been recently confirmed to follow the distribution in Eq. (\ref{pl}) \cite{juhasz2021}.  
In the model with long-range hopping, we expect for not too small $\alpha$ the states still to be (power-law) localized around two centers on the two sublattices and thus it makes sense to determine the maximum positions defined in Eq. (\ref{maxima}) and the distance between them.

The relevance of the distance between localization centers is that its distribution is closely related to the average entanglement entropy of a subsystem of contiguous sites.
In general, for a quantum system being in pure state $|\psi\rangle$, the entanglement entropy \cite{horodecki,amico,eisert2010,entanglement_review,laflorencie_rev} of a subsystem $A$ is defined as the von Neumann entropy of the reduced density matrix of $A$, $\rho_{A} = \Tr_{\overline{A}} | \psi \rangle \langle \psi |$: 
\be
S_A = -\Tr_A \rho_A \ln \rho_A.
\ee  
Here, $\Tr_{A}$ and $\Tr_{\overline{A}}$ stand for a partial trace over the subsystem and its environment, respectively. 
For the RSS, the entanglement entropy of a subsystem is easy to evaluate. Each occupied one-particle eigenstate which is shared between the subsystem and the environment gives a contribution of $\ln 2$. 
The average entanglement entropy of a block of $\ell$ contiguous sites in an infinite system can then be written as:  
\be 
\frac{S_{\ell}}{\ln 2}=\sum_{l<\ell}p_ll + \ell\sum_{l\ge \ell}p_l.
\label{Ssum}
\ee
Using the asymptotic distribution given in Eq. (\ref{pl}), the first term in Eq. (\ref{Ssum}) results in a leading logarithmic divergence of the average entanglement entropy \cite{refael}: $S_{\ell}\simeq \frac{\ln 2}{3}\ln\ell$.
In translationally invariant critical chains, the entanglement entropy is known to increase generally as $S_{\ell}\simeq \frac{c}{3}\ln\ell$ where $c$ is the central charge of the underlying conformal algebra \cite{holzhey,vidal,Calabrese_Cardy04}. 
In analogy to this, an effective central charge can be defined through the 
asymptotic relation $S_{\ell}\simeq\frac{c_{\rm eff}}{3}\ln\ell$, with 
$c_{\rm eff}=\ln 2$ for the short-range model \cite{refael}.
   
By general considerations one can show for SDRG schemes of one-dimensional models that the cumulative distribution $P_>(l)$ must have a tail $P_>(l)\simeq bl^{-1}$. 
Assuming a RSS with such a more general asymptotic distribution, the corresponding effective central charge determined by the first sum in Eq. (\ref{Ssum}) will be
\be
c_{\rm eff}=(3\ln 2)b.
\label{ceffb}
\ee
   
In fact, the relation between the average entanglement entropy and the cumulative distribution in a strictly localized random-singlet state goes beyond the equivalence of asymptotic prefactors. 
Defining a size-dependent effective central charge as a discrete derivative by $\ln\ell$ as 
\be  
c_{\rm eff}(\ell)=3\frac{S_{\ell+1}-S_{\ell}}{\ln(\ell+1)-\ln\ell},
\label{der0}
\ee
one obtains by using Eq. (\ref{Ssum}) 
$c_{\rm eff}(\ell)=3\ln2[\ln(\ell+1)-\ln\ell]^{-1}\sum_{l>\ell}p_l$.
Introducing a size-dependent effective prefactor $b(l)$ of the distribution as 
\be 
b(l)=lP_>(l),
\label{bl}
\ee
we have then for large $l$:
\be 
c_{\rm eff}(l)=(3\ln 2) b(l)\left[1+\frac{1}{2}l^{-1}+O(l^{-2})\right].
\ee
Nevertheless, numerical results obtained for the short-range model by exact diagonalization showed that this relationship is satisfied only for the asymptotic values in the limit $l\to\infty$ [Eq. (\ref{ceffb})], but not for the coefficients of the leading $O(1/l)$ corrections of $b(l)$ and $c_{\rm eff}(l)$, which must be due to the imperfect localization of states \cite{juhasz2021}.  

Thus we can see that if the ground state is a RSS, there must be a relation between the prefactor $b$ appearing in the distribution $P_>(l)$ of singlet lengths and $c_{\rm eff}$ appearing in the entanglement entropy. 
Computing both quantities for a model in an unknown state, we can then indirectly infer on whether this state is a RSS. One of our aims is to carry out this comparison for the long-range hopping model. 

The entanglement entropy of a fermion system can be efficiently computed from the eigenvalues of the correlation matrix $C_{ij}=\langle c_i^{\dagger}c_j\rangle$ restricted to the subsystem \cite{vidal,peschel}. The (symmetric) correlation matrix also inherits the bipartite structure of the model and, apart from the diagonal, its elements are non-zero only for indices of different parity: $C_{2i-1,2j}=C_{2j,2i-1}\equiv\overline{C}_{ij}=-\frac{1}{2}\sum_{k=1}^{L/2}\phi_k(i)\psi_k(j)$.  
For a subsystem of size $\ell$, the entanglement entropy can then be computed from the singular values of the $\ell/2\times\ell/2$ matrix $\overline{C}$.

\section{Strong-disorder renormalization group}
\label{sec:sdrg}

The idea of the SDRG approach developed originally for AF spin chains was later applied to random tight-binding models \cite{motrunich,melin,hoyos,junction}. 
The decimation rules of this method in the general case with both diagonal and off-diagonal disorder are known to coincide with the renormalization rules found by Aoki at zero energy, which exactly preserve the Green function of the renormalized system \cite{aoki,mg}.
The core of the SDRG procedure for the model defined in Eq. (\ref{hamilton}) is the following. 
Assume that there is a large coupling $t_{ij}$ in the system, compared to which all other couplings are negligible. 
The block of sites $n=2i-1$ and $m=2j$ has (in zeroth order) two half-filled (one-particle) states, $\psi_{\pm}(l)=\frac{1}{\sqrt{2}}(\delta_{nl}\pm\delta_{ml})$, with energies $\pm t_{ij}$. These states are then good approximations of those of the full system. Treating the rest of the system perturbatively, the block $nm$ is eliminated from the system and the original Hamiltonian is projected to an effective one in which hopping terms appear between pairs of sites connected to $n=2i-1$ and $m=2j$ in the second order of the perturbation with the amplitude:
\be 
\tilde t_{kl}=t_{kl}-\frac{t_{kj}t_{il}}{t_{ij}}.    
\label{rule}
\ee
Note that this elimination keeps the chiral symmetry since hopping terms within the sublattices and one-site (potential) terms are never generated. 
In the SDRG procedure, this step is iteratively applied to the block with the largest (in magnitude) coupling $\Omega=\max\{|t_{ij}|\}$, eliminating step by step pairs of high energy states and thereby reducing gradually the energy scale $\Omega$.   
The accuracy of this recursion depends on how the typical coupling ratio $t_{ij}/\Omega$ behaves as the renormalization proceeds. In the nearest-neighbor hopping model, the variance of the logarithmic couplings grows as $\sigma\sim n^{-1/2}$ as the fraction of active (i.e. non-decimated) sites $n$ decreases, thus typically $t_{ij}/\Omega\to 0$, ensuring the asymptotic exactness of the procedure \cite{fisherxx}. 
Instead of this infinite-disorder fixed point, in long-range models, the bare long-range couplings prevent the distribution of renormalized logarithmic couplings from an unbounded broadening, and the renormalization flow is attracted by a finite-disorder fixed point \cite{jki,moure,kettemann}. Here, the ratios $t_{ij}/\Omega$ are typically small but non-vanishing and the asymptotic exactness of the procedure is no longer guaranteed.  

We can see that, at each elimination step of the SDRG algorithm, all the remaining couplings have to be renormalized. For a sample of size $L$, the computational demand of determining the complete pairing structure by this algorithm is thus $O(L^3)$, which is much higher than that of the short-range model [$O(L\ln L)$].
We therefore also constructed a simpler SDRG scheme which has a linear (one-dimensional) structure and thereby has a lower, $O(L\ln L)$ computational demand, but still captures long-range hoppings at a minimal level.  
This is based on the observation that a perturbative correction (the second term on the r.h.s. in Eq. (\ref{rule}) is typically smaller (relative to $t_{kl}$) for farther sites. 
With a slight abuse of notation, we will denote the hopping amplitude between sites $n=2i-1$ and $m=2j$ by $t_{nm}$ rather than by $t_{ij}$, in the rest of this section, for a better clarity.  
Thus, considering a sequence of consecutive active sites labeled as $1$,$2$,$3$, and $4$, and assuming that $|t_{23}|$ is the maximal coupling so that site $2$ and $3$ are eliminated, then typically the most relevant correction is that added to the coupling between site $1$ and $4$. 
The main simplification of the minimal scheme is that renormalization is restricted only to the coupling connecting the left ($1$) and right ($4$) neighbor of the decimated pair, and other couplings are left unchanged. 
Furthermore, to obtain a one-dimensional SDRG scheme we must exclude decimations of pairs of farther-neighboring active sites. In the full SDRG scheme such decimations are atypical at large scales, so it makes a real restriction only at early stages of the procedure. 
These rules result in an effective one-dimensional scheme, in which two parameters, the couplings and distances between neighboring active sites are recorded only. The renormalization rule for eliminating sites $2$ and $3$ with $\Omega=|t_{23}|$ then reads as     
\beqn
\tilde t_{14}=w_{14}l_{14}^{-\alpha}-\frac{t_{12}t_{34}}{t_{23}} 
\label{rule2a} \\
\tilde l_{14}=l_{12}+l_{23}+l_{34}.
\label{rule2b}   
\eeqn
Note that the perturbative corrections to all other pairs of sites, which are neglected within this approximation, involve at least one coupling which connects non-neighboring active sites. 
Thus, in other words, in the minimal SDRG scheme only that perturbative term is kept which involves couplings between neighboring active sites.

\section{Conjectures based on the SDRG approach}
\label{sec:conjectures}

\subsection{Non-universality of the effective central charge}

Although the minimal SDRG scheme seems to be a crude approximation of the full scheme, we anticipate that it reproduces the prefactor $b$ (or $c_{\rm eff}$) obtained by the full scheme more and more accurately with increasing $\alpha$. 
Furthermore, this simplified formulation of the renormalization is also instructive for describing the general characteristics of the renormalization flow of the long-range hopping model. 
Let us therefore inspect the decimation rules formulated in Eqs. (\ref{rule2a}-\ref{rule2b}).
Dropping the first term on the r.h.s. of Eq. (\ref{rule2a}), which describes long-range hopping, we recover the decimation rule of the nearest-neighbor model. 
Here, it is well-known that $c_{\rm eff}=\ln 2$, or, equivalently $b=\frac{1}{3}$.

Let us now consider the opposite case and omit for a while the second term in Eq. (\ref{rule2a}), which describes the effective hopping through the decimated sites. In the limiting case, when the distribution of $w$ is narrow, the prefactor $b$ can be calculated from earlier exact results. Obviously, in this limit, the amplitudes are perfectly correlated with the bond lengths, and the largest amplitude is associated with the shortest bond length. Then, in the SDRG procedure, always the actually shortest  bond is eliminated. This scheme is formally identical with a simplified model describing the coarsening of the one-dimensional Glauber-Ising model started from a random initial state at zero temperature, for which the distribution of distances between domain walls at late times have been exactly calculated \cite{nagai,bray,rutenberg}. The results we need for the calculation of $b$ are the following. i) The distribution of lengths $l$ has a scaling property $f_{\mathcal{L}}(l)=\mathcal{L}^{-1}F(l/\mathcal{L})$ for $\mathcal{L}\to\infty$, where $\mathcal{L}=\min_n\{l_n\}$. ii) The functional value of the scaling function at the edge of the distribution is $F(1)=\frac{1}{2}$. iii) The minimal length is related to the average as 
$\mathcal{L}=\frac{1}{2}e^{-\gamma_E}\overline{l}$, where $\gamma_E=0.577215\dots$ is Euler's constant.        
When the minimal length is changed by $d\mathcal{L}$ in the course of the SDRG procedure, then singlets of length $\mathcal{L}$ are produced, and their probability density changes by $dp(\mathcal{L})=2n_{\mathcal{L}}f(\mathcal{L})d\mathcal{L}$, where $n_{\mathcal{L}}=1/\overline{l}$ is the fraction of remaining bonds. 
This results in $P_>(l)\simeq \frac{1}{2}e^{-\gamma_E}l^{-1}$ for the tail of the distribution, so in the limiting case of narrow disorder distribution we have a prefactor:
\be 
b=\frac{1}{2}e^{-\gamma_E}=0.280729\dots
\label{bmin}
\ee
and a corresponding effective central charge 
$c_{\rm eff}=(3\ln 2)b=0.583761\dots$, independently of $\alpha$. 
Note that this limiting value is obtained also for an arbitrary distribution of $w$ in the limit $\alpha\to\infty$.   
If we now relax the requirement on the narrowness of the distribution of $w$ [but still omitting the second term in Eq. (\ref{rule2a})], then we have $\ln\tilde t_{14}=\ln w_{14}-\alpha\ln l_{14}$, and the random term $\ln w_{14}$ will break the perfect correlations between $t$ and $l$. Since $f_{\mathcal{L}}(l)$ still has the scaling property, the distribution of $\ln l$ will not broaden during the SDRG procedure, consequently the random term $\ln w$ cannot be neglected asymptotically and, according to our numerical results (not shown), leads to non-universal prefactors, which depend on the distribution of $w$. 
For a disorder distribution with a finite width, the correlation between $t$ and $l$ is not perfect, hence not always the shortest bonds are eliminated and, as a consequence, the prefactor $b$ will be higher than that given in Eq. (\ref{bmin}). For this reason, we conjecture that this value is a lower bound for the values that can be obtained for random-singlet states by the SDRG method.

Now returning to the complete minimal scheme specified by Eqs. (\ref{rule2a}-\ref{rule2b}), we expect the presence of the second term in Eq. (\ref{rule2a}), which is typically in the same order of magnitude as the first one, at most to temper the non-universality, i.e. to make the dependence of $b$ on the disorder distribution weaker, but not to be able to eliminate it completely. This is confirmed by numerical results obtained with different disorder distributions, see later.  
Based on these observations on the minimal scheme, we expect the effective central charge to be non-universal, i.e. dependent on $\alpha$ and the disorder distribution within the full SDRG approximation, as well as in the original model.

\subsection{The effect of the short-range fixed point}

According to the general scenario of the critical behavior of homogeneous systems with long-range interactions, the critical behavior of the short-range system is recovered i.e. long-range interactions become irrelevant if $\alpha$ exceeds a finite, model-dependent threshold \cite{nickel,sak}. 
Disordered systems, the short-range variant of which displays an infinite-randomness critical behavior are, however, much different in this respect. 
In such short-range models, the relationship between the energy scale $\epsilon$ and the length scale  $\xi$ is extremely anisotropic, having the form $|\ln\epsilon| \sim \sqrt{\xi}$ \cite{fisherxx}. 
In the long-range variant, the bare long-range couplings prevent the energy scale from a rapid decrease with the length scale and the dynamical relationship is dictated by the distance-dependence of long-range couplings to be $\epsilon\sim \xi^{-\alpha}$. This relationship manifests itself e.g. in the finite-size scaling of the energy gap, $\epsilon_L\sim L^{-\alpha}$ \cite{jki,moure}.
Thus, the dynamical relationship of the short-range hopping model is not recovered by increasing $\alpha$, no matter how large $\alpha$ is. 
This suggests that characteristics of the ground state like the effective central charge may also be different from that of the short-range model even for an arbitrarily large $\alpha$.  

There is, however, a difficulty with studying the large $\alpha$ regime: the effect of long-range hopping appears only at large scales. 
To qualitatively describe this crossover we invoke that, according to the solution of the SDRG flow equation in the short-range model, the effective hopping amplitudes between sites in a distance $\xi$ scale typically as $\tilde t\sim \exp(-C\sqrt{\xi})$, where $C$ is an $O(1)$ random variable \cite{fisherxx}. In the long-range model with a large $\alpha$, the bare long-range couplings may then be negligible compared to the effective couplings realized by a chain of intervening nearest-neighbor bonds at small scales, and they start to be relevant only beyond a crossover scale $\xi_*$, which is naively determined by comparing the two kinds of contributions:
\be 
e^{-C\sqrt{\xi_*}}\sim \xi_*^{-\alpha}. 
\ee
This gives for the crossover length scale 
\be 
\xi_*\sim \left[\frac{\alpha}{C}\ln\left(\frac{\alpha}{C}\right)\right]^2.
\label{xi}
\ee
Within this scale, the system is thus expected to behave as a short-range system and a new type of critical behavior is expected to appear only well beyond $\xi_*$, which can be rather large for a large $\alpha$.

\section{Numerical results}
\label{sec:numerical}

We used in the numerical calculations different types of arbitrarily chosen distributions of $w_{ij}$ in order to test a possible non-universality of the effective central charge. 
One of them is a continuous (uniform) distribution with the support $[a,a+1]$, which we used with $a=-\frac{1}{2}$, where the signs of hopping amplitudes can be both positive and negative and with $a=0$, where all amplitudes are positive. 
The other one is a discrete bimodal distribution, in which $w_{ij}$ can be either $1$ or $1/3$ with equal probabilities.
The calculations were performed for various values of the decay exponent, mainly in the range $\alpha\in [1,5]$. 
The eigenvalue problem of the model was solved by applying singular-value decomposition as described in section \ref{sec:model}. In the following this method will be referred to as exact diagonalization (ED). By ED, we calculated the distribution of the distances between localization centers, and the corresponding effective prefactor $b(l)$ defined in Eq. (\ref{bl}). The system size went up to $L=16384$ for smaller values of $\alpha$, otherwise it was limited by numerical precision problems of the numerical routine. The number of random samples was $10^4$ for $L=16384$, and $10^5$ for smaller sizes. 
We also computed the average entanglement entropy $S_{L}$ of blocks of size $\ell=L/2$ for a series of system sizes $L=16,32,64,\dots, 8192$, and determined the asymptotic dependence on the block size by finite-size scaling.
The number of random samples was typically $10^6$ (for the largest size only $5\times 10^4$) and an additional average was also performed for $16$ different positions of the block in the periodic system.  
As we observed an overall logarithmic dependence of the entanglement entropy on the block size with strong corrections, we calculated size-dependent effective  central charges from adjacent data points: 
\be 
c_{\rm eff}(L)=3\frac{S_{2L}-S_{L}}{\ln 2}.
\label{der}
\ee  
For the relatively time-consuming full SDRG method, the maximal system size we could consider was $L=4096$. Here the distribution of singlet lengths was calculated in $10^5$ random samples. Within this approach, the average entanglement entropy can readily be obtained through Eq. (\ref{Ssum}).  
The much more efficient minimal SDRG scheme was performed in $10^6$ random samples of maximal size $L=10^6$.    

\subsection{Dependence on $\alpha$}

\begin{figure}
\begin{center}
\includegraphics[width=80mm, angle=0]{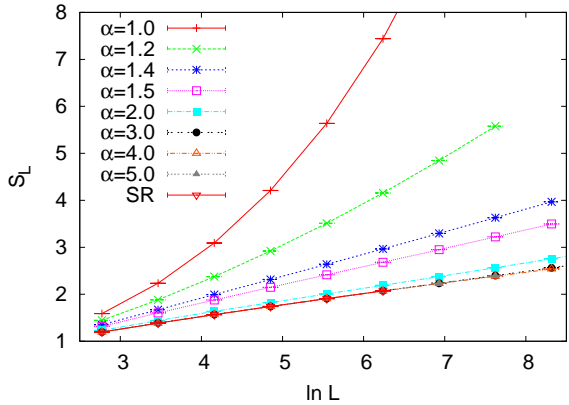}
\end{center} 
\caption{
\label{fig_sl} The average half-system entanglement entropy plotted against the logarithm of the system size for different values of the decay exponent $\alpha$. As a comparison, data obtained for the short-range model with nearest-neighbor hopping (SR) is also shown. 
 Note that the data for $\alpha=4,5$, and SR can hardly be distinguished from each other.
The data were obtained by exact diagonalization.
}
\end{figure}

\begin{figure}
\begin{center}
\includegraphics[width=80mm, angle=0]{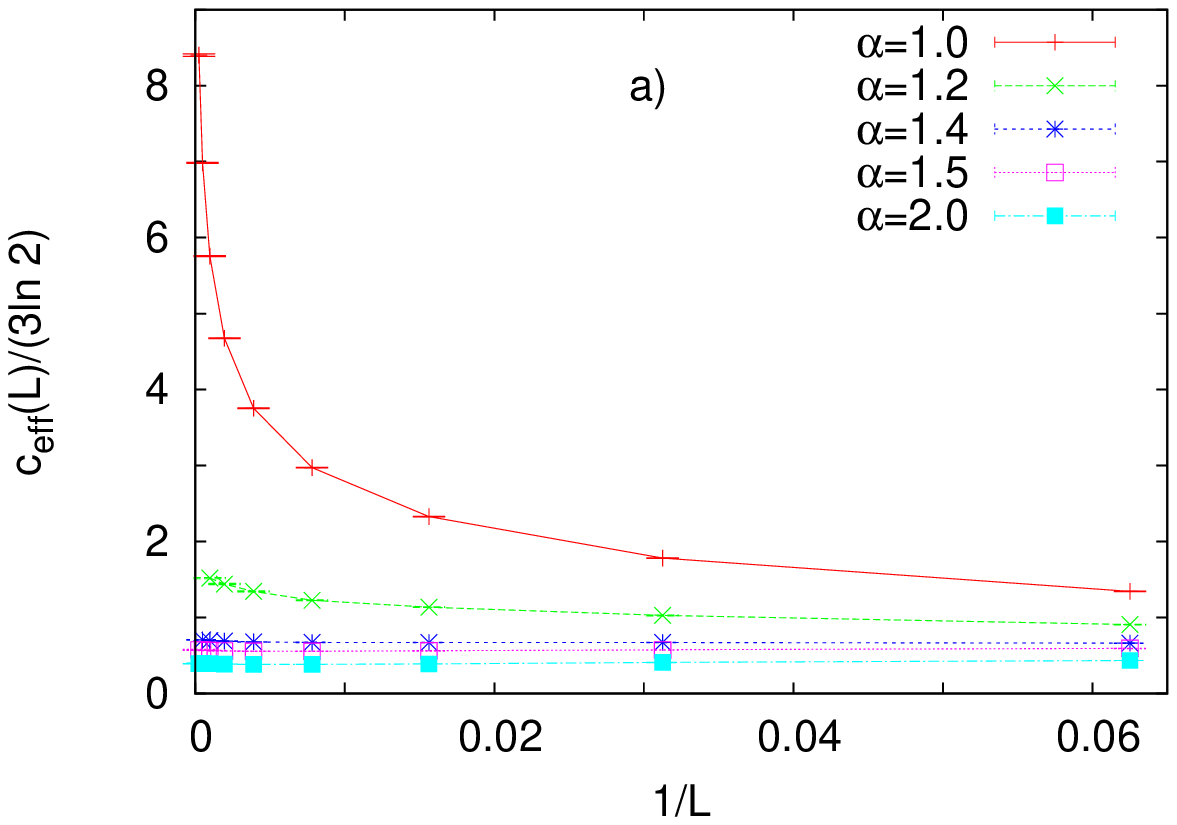}
\includegraphics[width=80mm, angle=0]{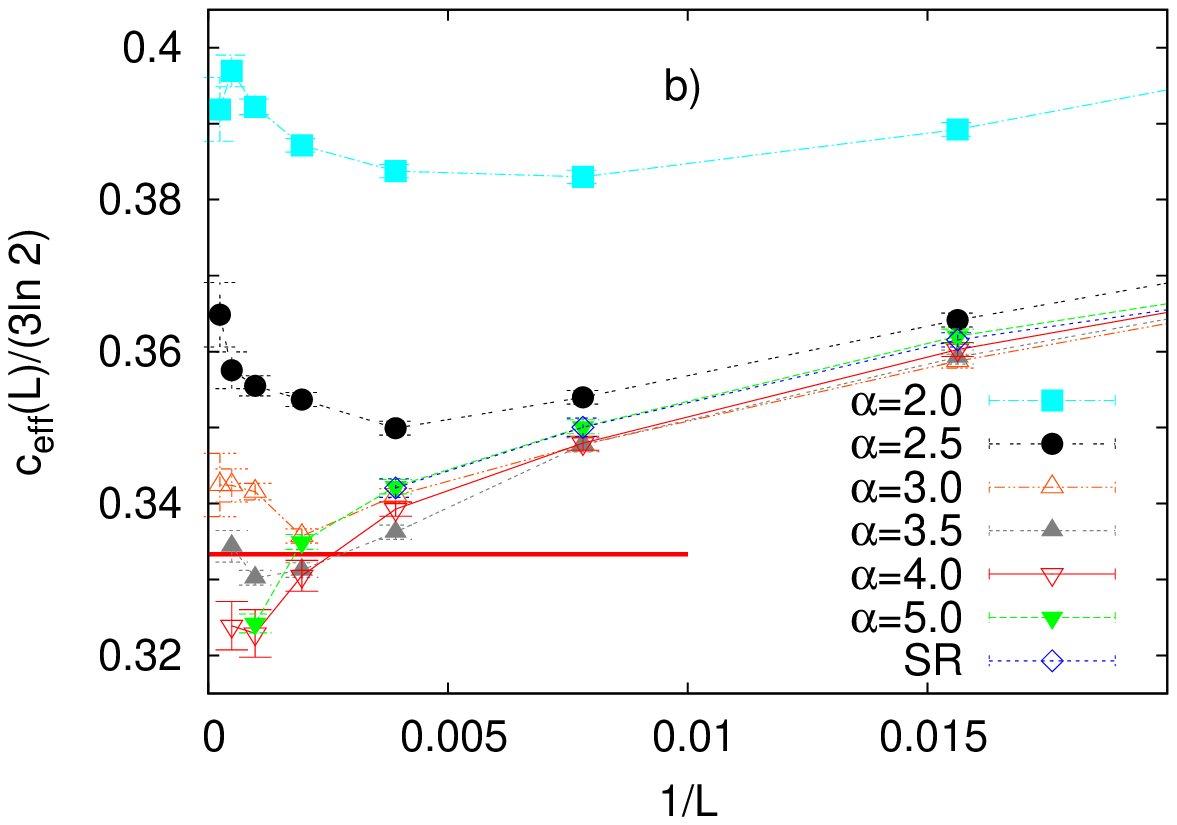}
\end{center} 
\caption{
\label{fig_ceff_ed}
The size-dependent effective central charge plotted against $1/L$ for different values of $\alpha$. The red horizontal line at $c_{\rm eff}/(3\ln 2)=1/3$ indicates the asymptotic value of the short-range model. The data were obtained by exact diagonalization.
}
\end{figure}

The numerical data presented in this and the next subsection were obtained with the uniform distribution of the random factors $w_{ij}$ with $a=-\frac{1}{2}$. 
Other disorder distributions are probed in subsection \ref{subsec:randomness}.
The average half-system entanglement entropy obtained by ED is shown as a function of the system size in Fig. \ref{fig_sl}. 
As can be seen, the dependence on $L$ is logarithmic in the extensive regime $\alpha>1$ and the data for large $\alpha$ hardly differ from those of the model with nearest-neighbor hopping.  
The size-dependent effective central charges are shown in Fig. \ref{fig_ceff_ed}. Since, for the short-range model the leading correction of $c_{\rm eff}(L)$ has been found to be $O(1/L)$ \cite{juhasz2021}, we plotted the data against $1/L$. 
At the boundary of the extensive regime, $\alpha=1$, $c_{\rm eff}(L)$ seems to diverge with increasing $L$, signaling a super-logarithmic divergence of the entanglement entropy, in accordance with the results of Ref. \cite{roy}. For $\alpha>1$, the data suggest a convergence of $c_{\rm eff}(L)$ to finite, $\alpha$-dependent limiting values in the large $L$ limit. A variation of the data with $\alpha$ can be clearly observed, which becomes weaker and weaker with increasing $\alpha$, suggesting a convergence of the asymptotic value $c_{\rm eff}$ for $\alpha\to\infty$ to a limit which is different from the central charge $c_{\rm eff}/(3\ln 2)=1/3$ of the short-range model.  
As opposed to the short-range model, the dependence of $c_{\rm eff}(L)$ on $L$ is non-monotonic for larger values of $\alpha$: after an initial decrease, $c_{\rm eff}(L)$ goes through a minimum and subsequently increases with $L$. 
Here, in the initial (small size) regime, an approach toward the short-range value $1/3$ (shown by a horizontal line in Fig. \ref{fig_ceff_ed}b) can be observed which persists for larger sizes with increasing $\alpha$. This transient regime is then followed by a (non-monotonic) crossover to the true asymptotic value.  

\begin{figure*}
\includegraphics[width=80mm, angle=0]{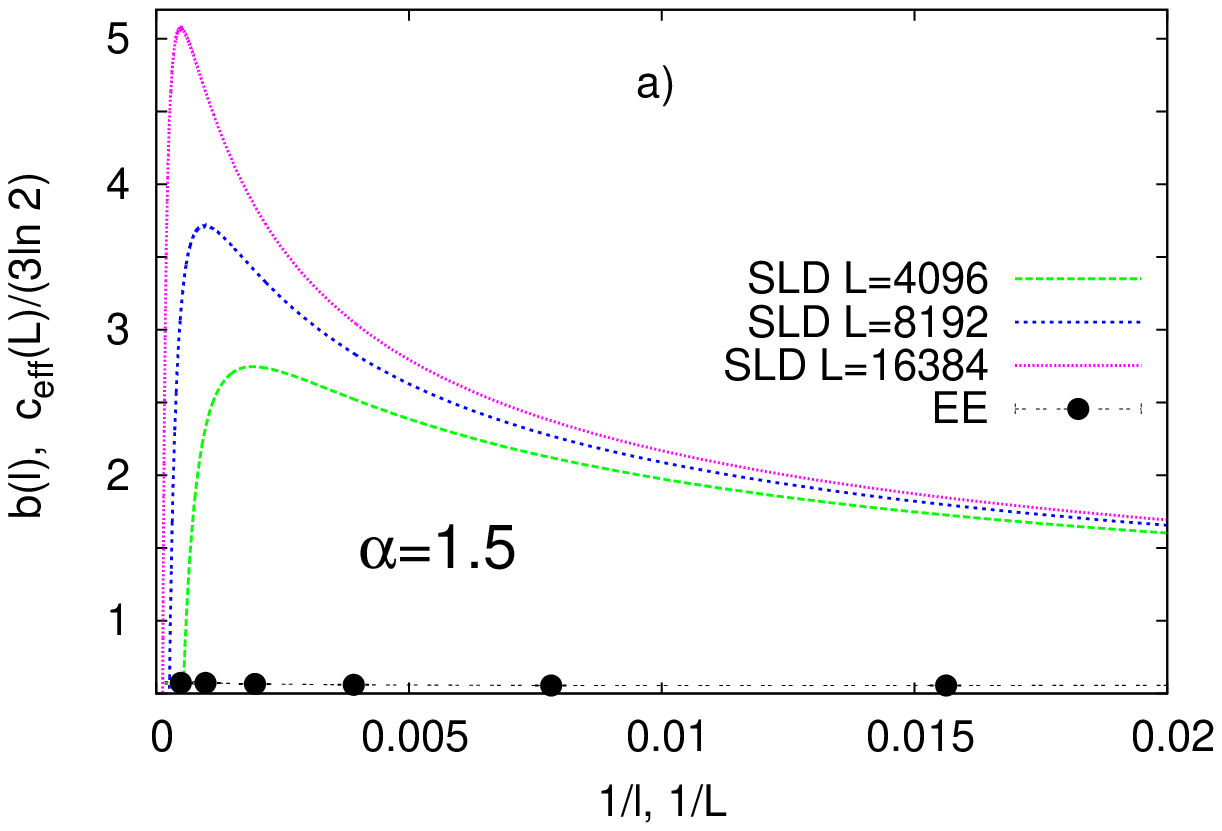}
\includegraphics[width=80mm, angle=0]{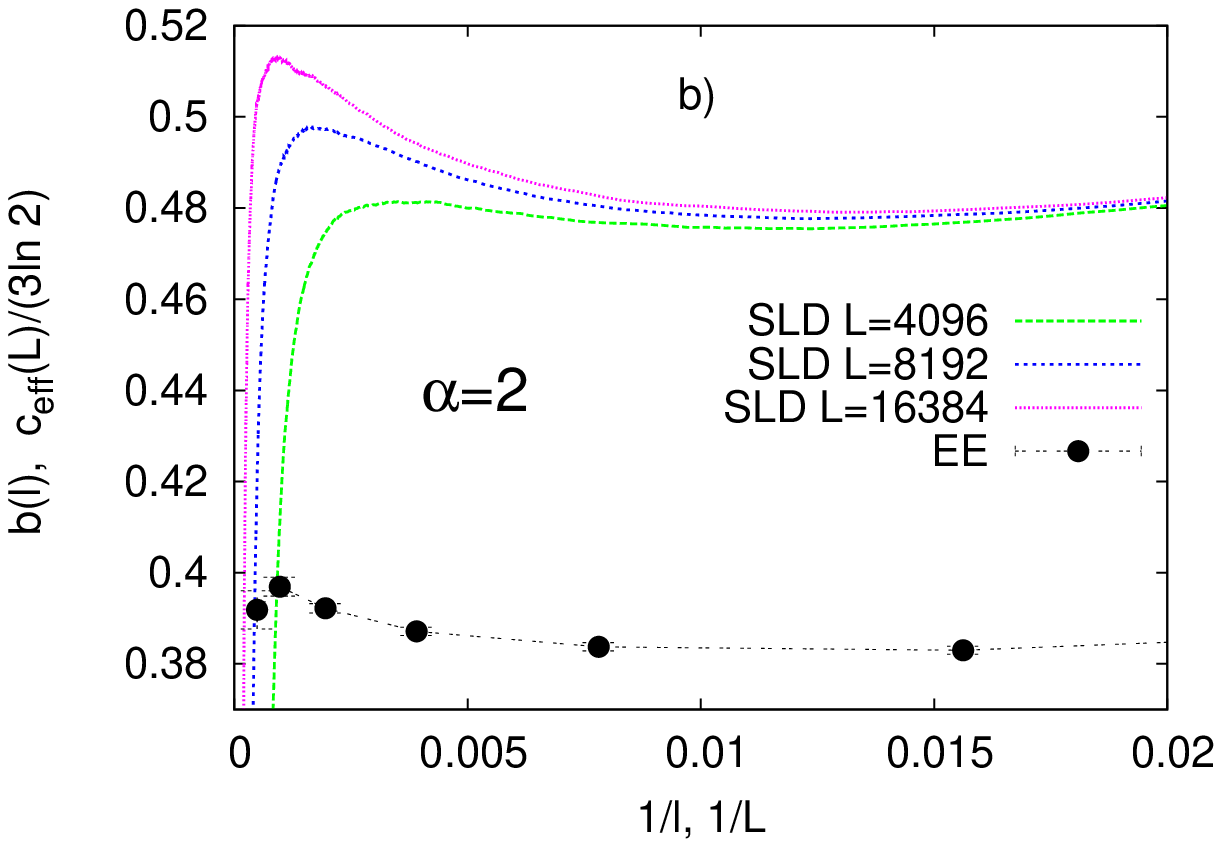}
\includegraphics[width=80mm, angle=0]{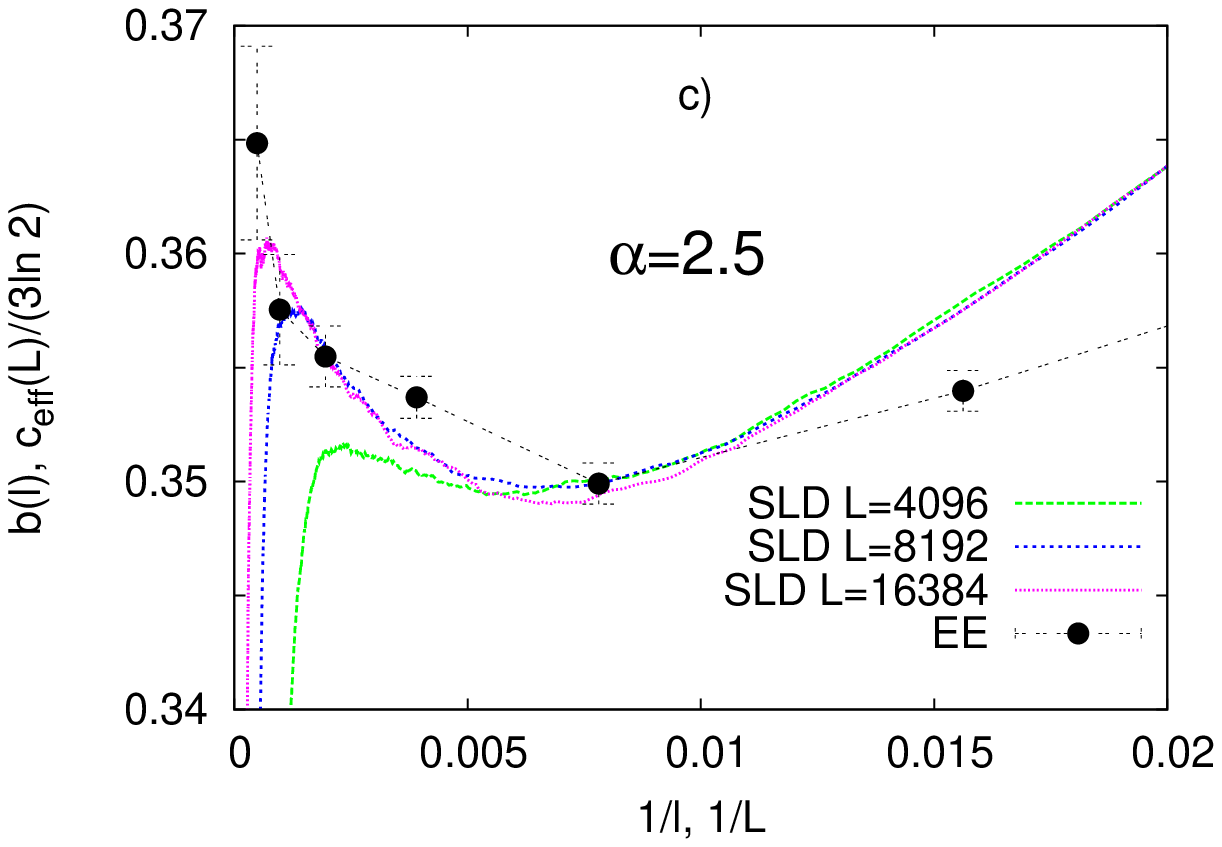}
\includegraphics[width=80mm, angle=0]{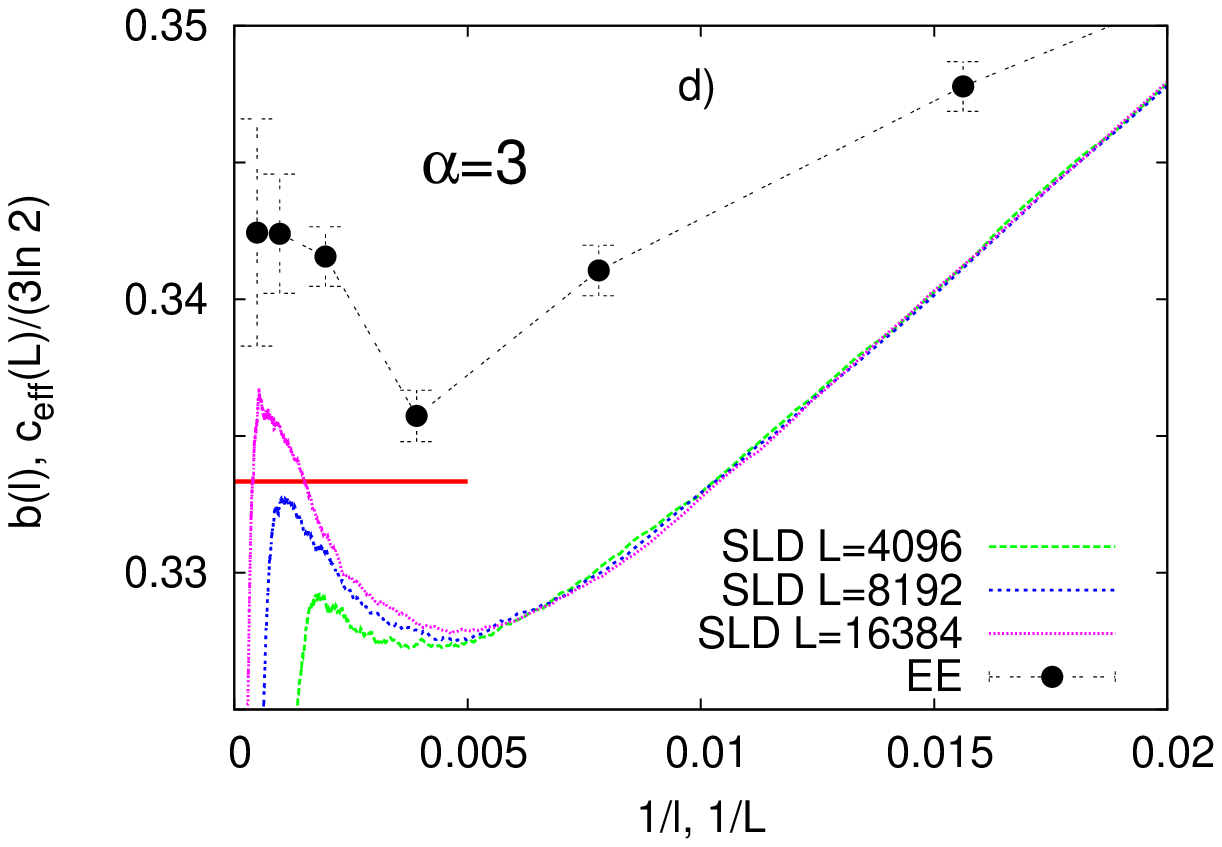}
\includegraphics[width=80mm, angle=0]{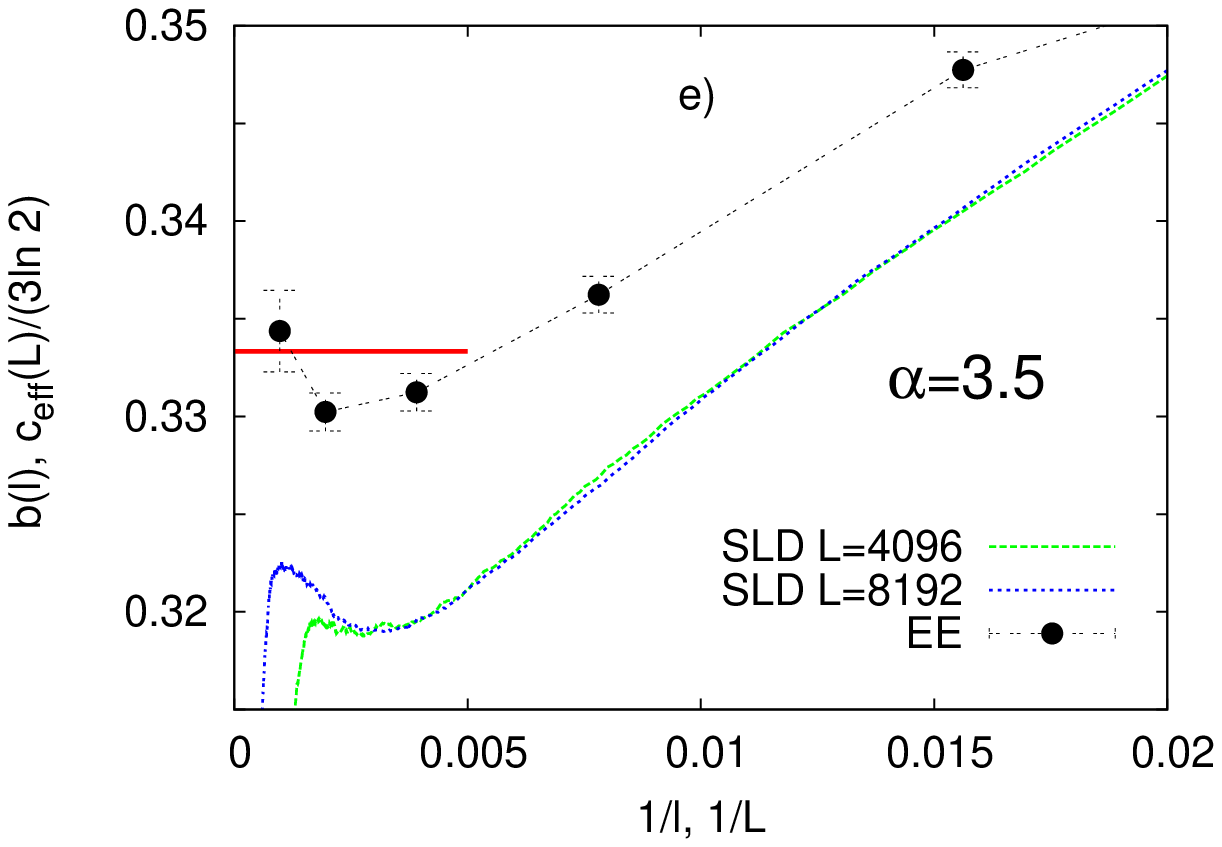}
\caption{
\label{fig_ceff_b} 
The effective prefactor $b(l)$ of singlet-length distribution (SLD) and the size dependent effective central charge (EE), $c_{\rm eff}(L)/(3\ln 2)$, plotted in the same figure against $1/l$ and $1/L$, respectively, for $\alpha=1.5$ (a), $\alpha=2$ (b), $\alpha=2.5$ (c), $\alpha=3$ (d), and $\alpha=3.5$ (e). The red horizontal line at $c_{\rm eff}/(3\ln 2)=1/3$ indicates the asymptotic value of the short-range model. The sharp cutoff in the SLD data at $l\sim L$ is a finite-size effect.
}
\end{figure*}

Next, we compare the effective central charge obtained from the entanglement entropy to the prefactor $b$ appearing in the the asymptotic form of the singlet-length distribution. For a RSS, the relationship $c_{\rm eff}=(3\ln 2)b$ must hold. 
The effective prefactor $b(l)=lP_>(l)$, which tends to $b$ in an infinite system in the limit $l\to\infty$ is plotted together with $c_{\rm eff}(L)/(3\ln 2)$ for different values of $\alpha$ in Fig. \ref{fig_ceff_b}.
As can be seen for $\alpha\le 2$, $b(l)$ seems to tend to a significantly higher value than $c_{\rm eff}(L)/(3\ln 2)$ in the large size limit; the relationship characteristic for a RSS is thus not fulfilled. 
At $\alpha=2.5$ the limiting values of the two quantities seem to be compatible. For $\alpha=3$ and $\alpha=3.5$, the finite-size effective central charges become higher than $b(l)$. Since for larger $\alpha$ the crossover region together with the minimum position of the curves is shifted to larger sizes, the steeply raising regime of the curve $b(l)$ beyond the minimum is strongly affected by the finite-size of the system, therefore a reliable estimate on the limiting value $b$ and a clear conclusion on the validity of the relationship $c_{\rm eff}=(3\ln 2)b$ cannot be given. 

\subsection{Comparison of different methods}

\begin{figure*}
\includegraphics[width=80mm, angle=0]{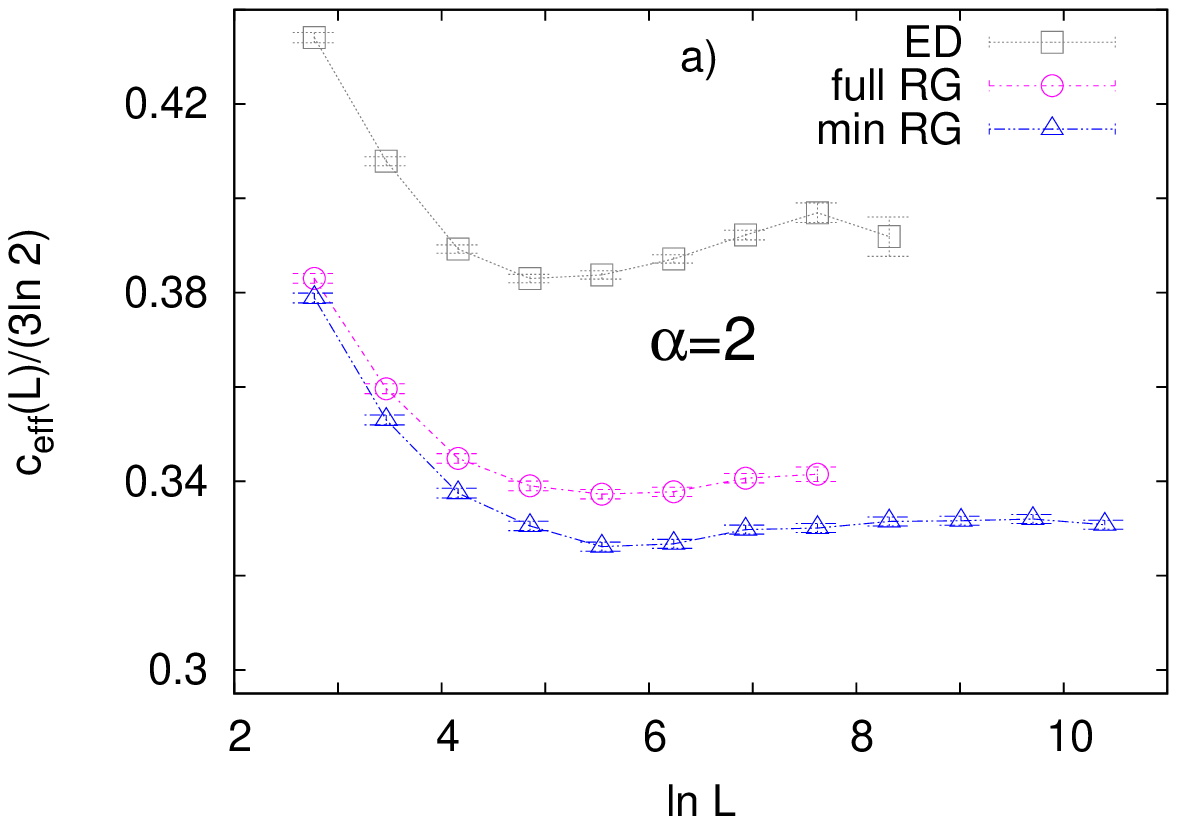}
\includegraphics[width=80mm, angle=0]{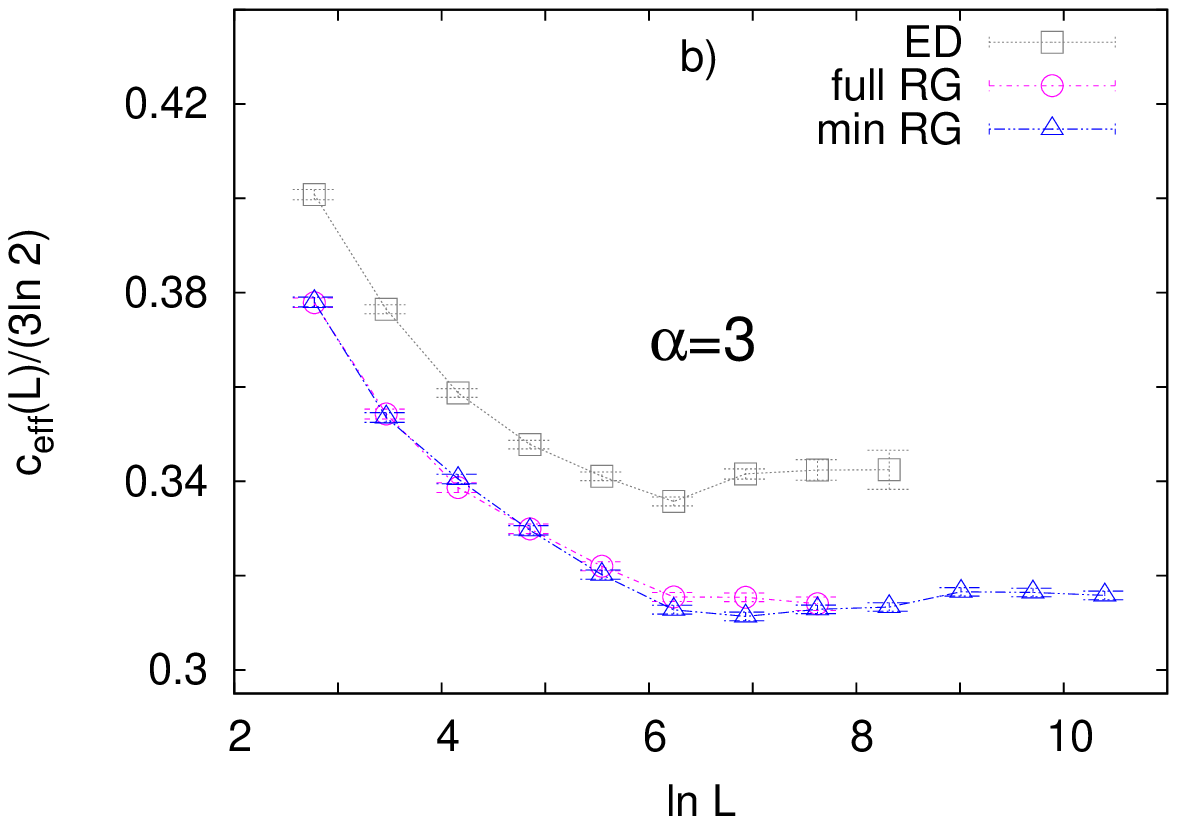}
\includegraphics[width=80mm, angle=0]{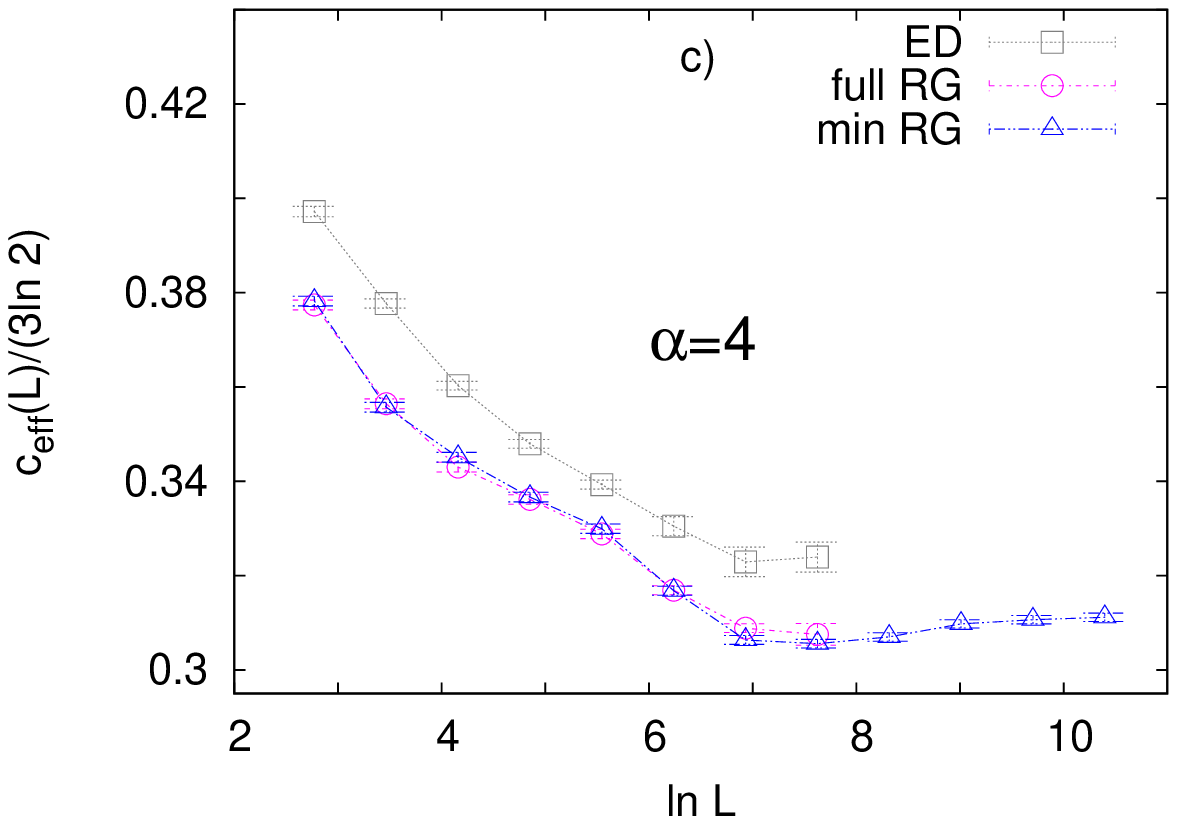}
\includegraphics[width=80mm, angle=0]{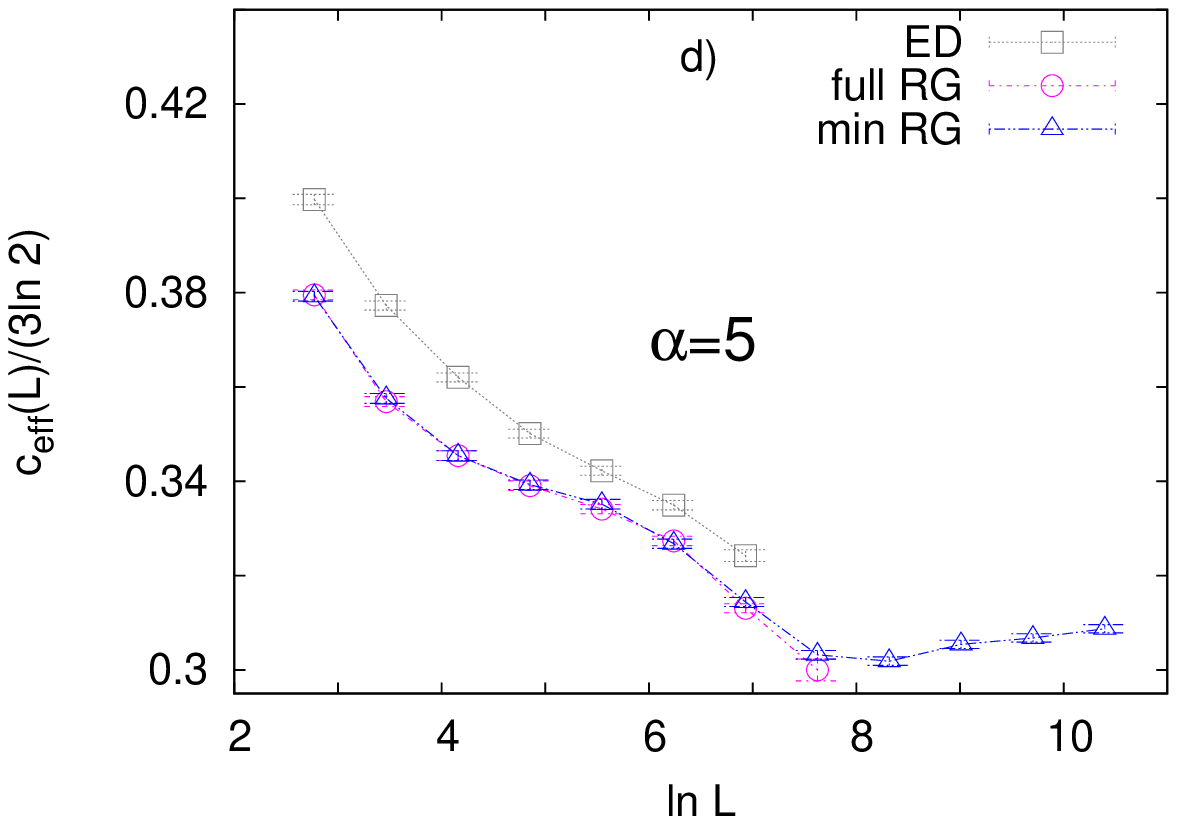}
\caption{
\label{fig_comp} 
The size dependent effective central charge as a function of $\ln L$ obtained by exact diagonalization (ED), the full SDRG method (full RG), and the minimal SDRG scheme (min RG) for $\alpha=2$ (a), $\alpha=3$ (b), $\alpha=4$ (c), and $\alpha=5$ (d). 
}
\end{figure*}

Next we check the validity of the SDRG method for the calculation of $c_{\rm eff}$. The finite-size effective central charges obtained by the full SDRG scheme and by the minimal SDRG scheme are compared to the ED results in Fig. \ref{fig_comp} for various $\alpha$.  
As can be seen, the variation of $c_{\rm eff}(L)$ with the system size is qualitatively similar by all methods, but the SDRG methods give systematically lower values than ED. Nevertheless, the deviation becomes smaller with increasing $\alpha$. One can also observe that the minimal SDRG scheme is a better and better approximation of the full scheme with increasing $\alpha$, and the difference in $c_{\rm eff}(L)$ obtained by the two methods is much smaller than the deviation from ED data. 

\begin{figure*}
\begin{center}
\includegraphics[width=80mm, angle=0]{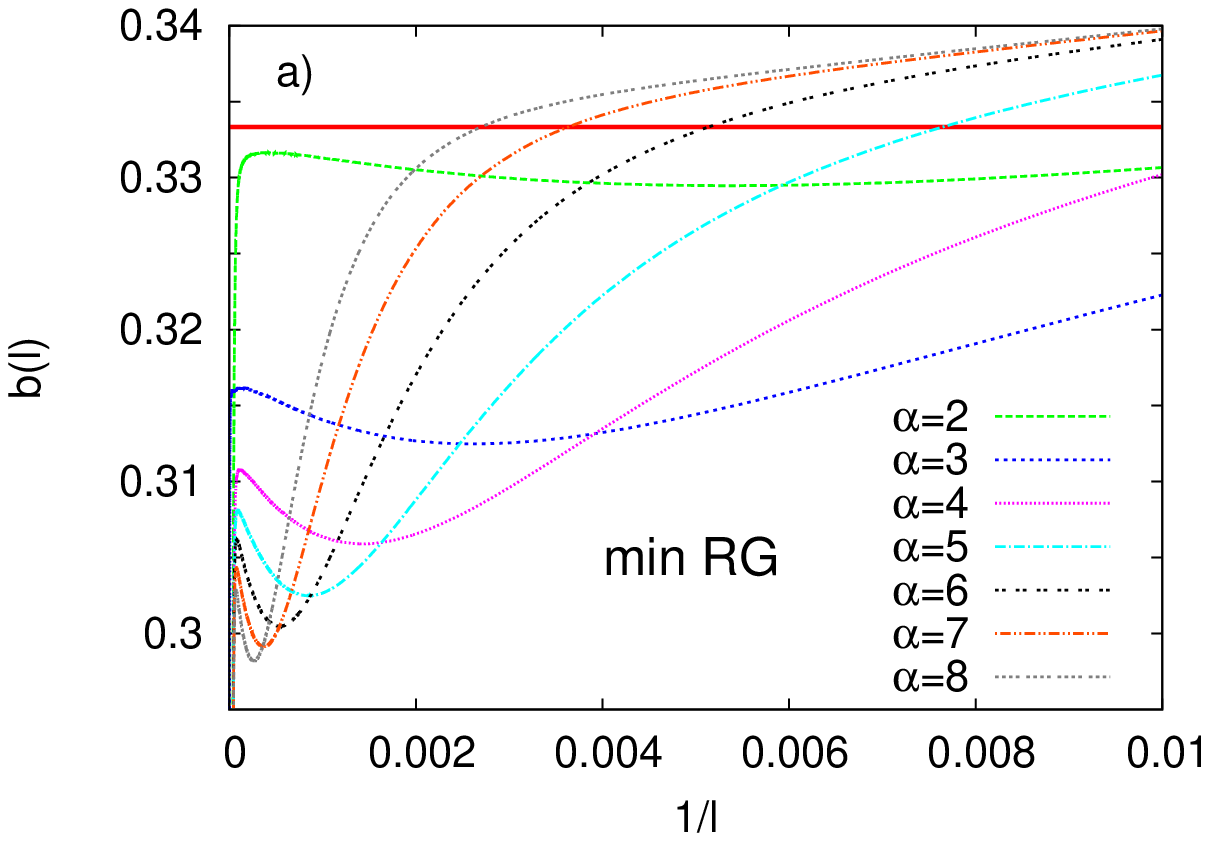}
\includegraphics[width=80mm, angle=0]{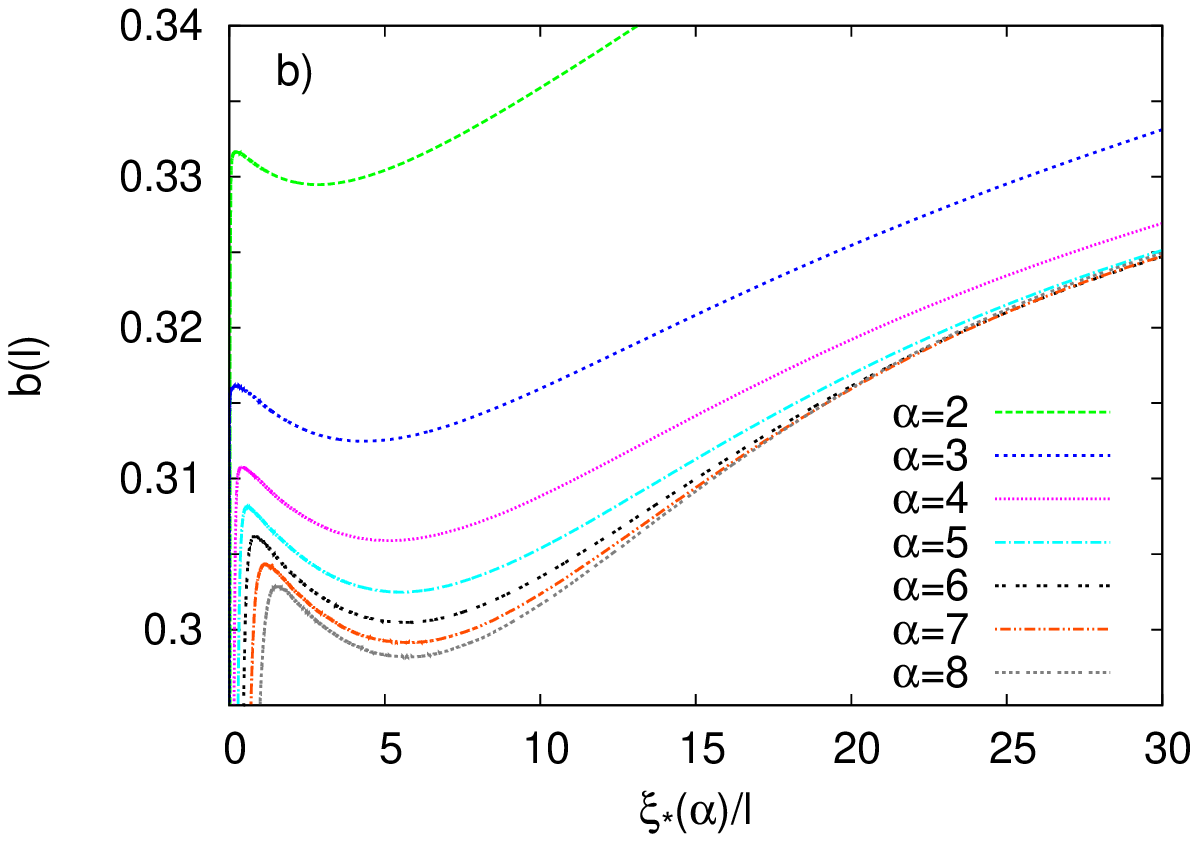}
\end{center} 
\caption{
\label{fig_minrg} 
The effective prefactor $b(l)$ of the singlet-length distribution for various $\alpha$, plotted against $1/l$ (a) and $\xi_*(\alpha)/l$ (b). Here, $\xi_*(\alpha)$ is the $\alpha$ dependent crossover length scale given in Eq. (\ref{xi}), evaluated with $C=0.2$. The data were obtained by the minimal SDRG scheme in systems of size $L=10^5$. The red horizontal line at $1/3$ indicates the asymptotic prefactor of the short-range model. 
}
\end{figure*}

Finally, we present the effective prefactor $b(l)$ of the singlet-length distribution, obtained by the minimal SDRG scheme for several $\alpha$ in Fig. \ref{fig_minrg}. The effect of the short-range fixed point can be clearly seen in the initial tendency of the data to $1/3$ followed by a crossover to the true (long-range) fixed point, which is shifted to larger and larger scales for increasing $\alpha$.
Plotting the data against $\xi_*(\alpha)/l$ with the crossover length scale $\xi_*(\alpha)$ in Eq. (\ref{xi}), as was done in Fig. \ref{fig_minrg}b, the shift of the crossover region is roughly compensated, confirming the validity of the simple reasoning resulted in Eq. (\ref{xi}).    

\subsection{Comparison of different disorder distributions}
\label{subsec:randomness}
\begin{figure*}
\includegraphics[width=80mm, angle=0]{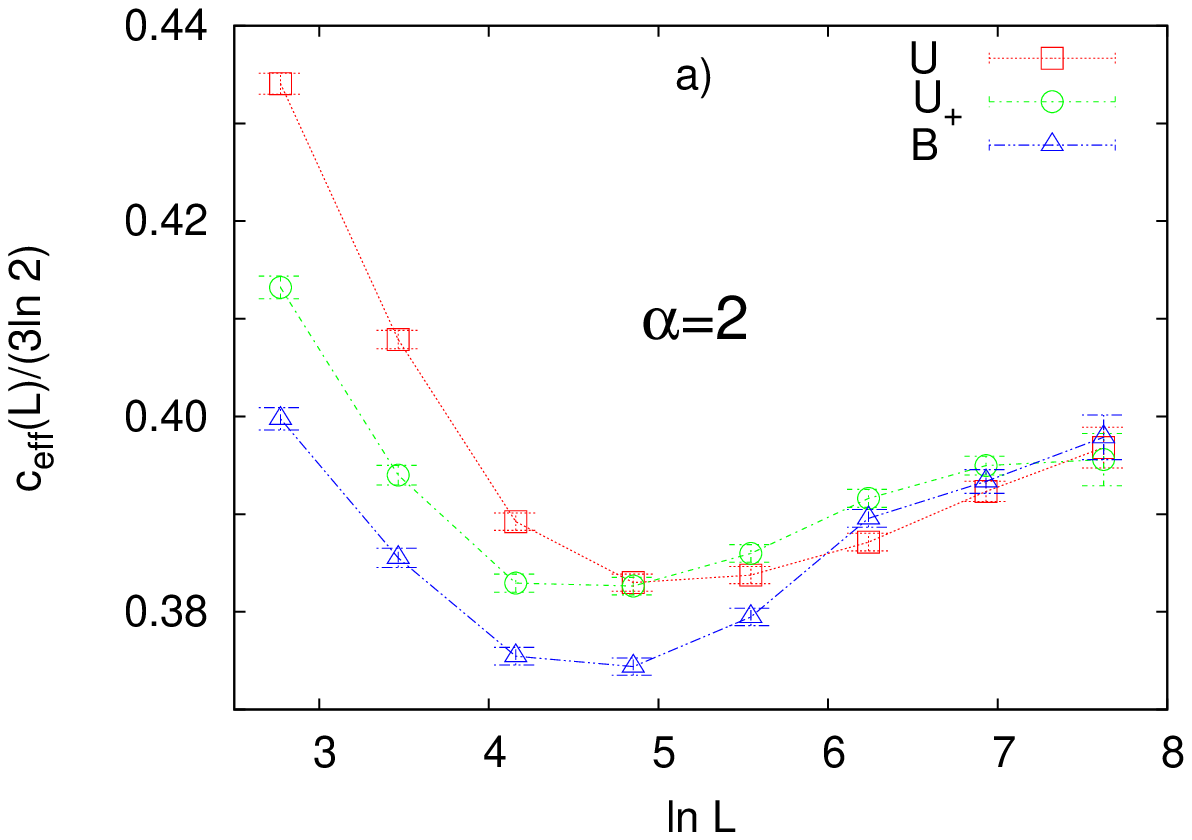}
\includegraphics[width=80mm, angle=0]{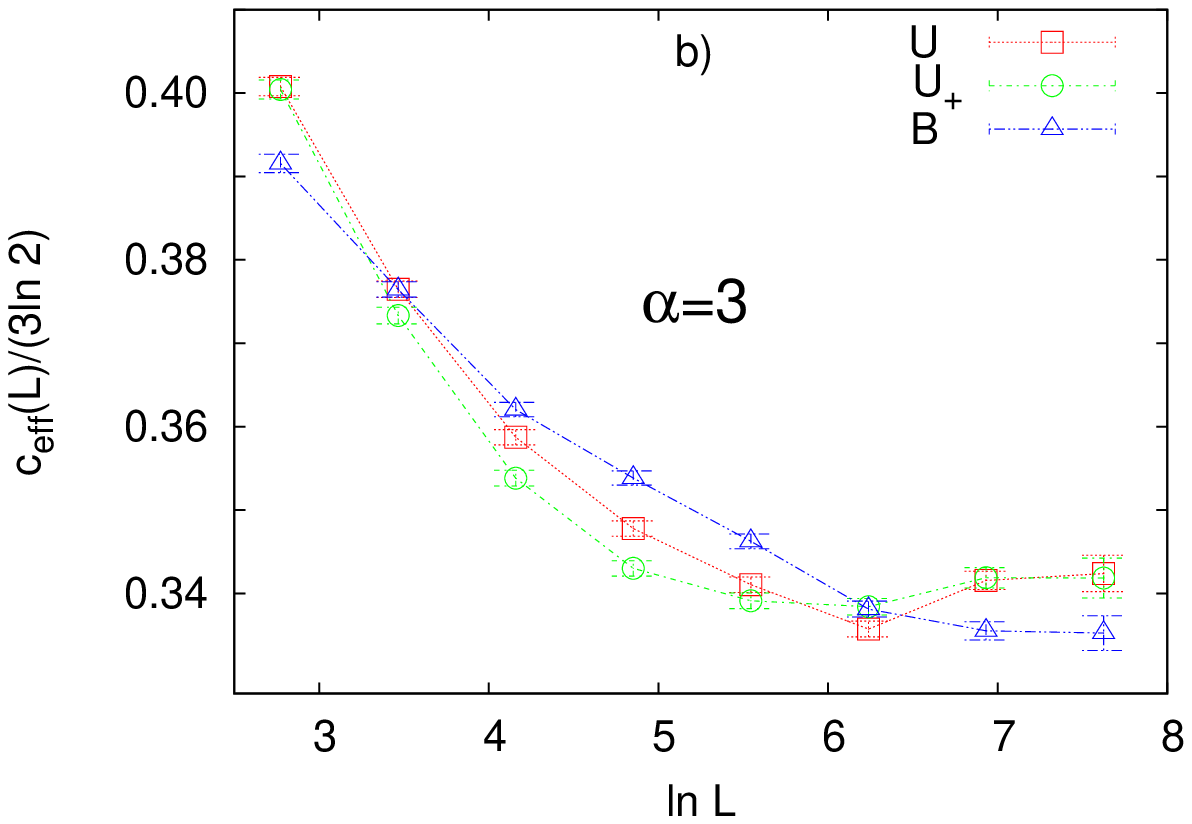}
\caption{
\label{fig_univ} 
The size dependent effective central charge as a function of $\ln L$, for uniform disorder distribution with $a=-\frac{1}{2}$ ($U$), with $a=0$ ($U_+$), and with the binary distribution ($B$) specified in the text. The data were obtained by ED for $\alpha=2$ (a) and $\alpha=3$ (b). 
}
\end{figure*}

Last, we address the question of universality with respect to the distribution of the random factors $w_{ij}$. 
The finite-size data obtained by ED are shown for three different distributions in Fig. \ref{fig_univ}. 
For $\alpha=2$, the differences between the data for small $L$ decrease with increasing $L$, down to the error of estimates. 
For $\alpha=3$, the data obtained with the binary distribution seem to be significantly below the other two at the largest sizes but the asymptotical region may not yet be reached here. 
The full SDRG method, owing to the limited system sizes is even less capable of pointing out a non-universality in the asymptotic values, see the effective prefactors $b(l)$ in Fig. \ref{fig_univ}.     
The only method by which this question is accessible is the minimal SDRG scheme. As can be seen for $\alpha=3$ in the inset of Fig. \ref{fig_univ}, the curves seem to tend to  different asymptotic values. The non-universality is, however, weak compared to the variation with $\alpha$, and is below the precision accessible by ED.    
\begin{figure}
\begin{center}
\includegraphics[width=80mm, angle=0]{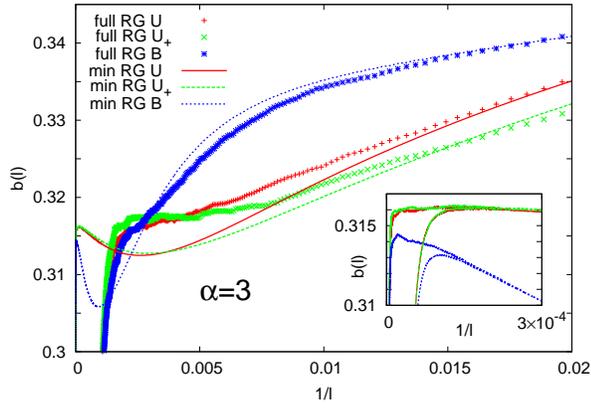}
\end{center} 
\caption{
\label{fig_univ_rg} 
The effective prefactor of the singlet-length distribution obtained with the full and the minimal SDRG scheme for $\alpha=3$ and for three different disorder distributions described in the caption of Fig. \ref{fig_univ}. The full and the minimal SDRG scheme were performed for system sizes $L=4096$ and $L=10^6$, respectively. The inset shows data obtained by the minimal SDRG scheme for $L=10^5$ (thin lines) and $L=10^6$ (thick lines) in the large scale regime.
}
\end{figure}

\section{Discussion}
\label{sec:discussion}

In this paper, we considered a system of free fermions on a one-dimensional lattice with power-law decaying, chiral symmetric, random hopping, and studied the question whether the concept of random-singlet state formulated originally for short-range models is valid for this model. 
For this purpose, we studied the effective central charge characterizing the logarithmic divergence of the entanglement entropy, the asymptotic value of which 
 is known to be universal and correctly reproduced by the SDRG method for the nearest-neighbor model.
We carried out two kinds of consistency checks. First, we compared the effective central charge to the prefactor appearing in the distribution of the distance between localization centers on odd and even sublattices. Second, the effective central charge obtained by ED was compared with that obtained by the SDRG method naively applied to the model, which produces a pair-localized ground state through an iterative scheme. To make large systems accessible, we constructed a minimal  SDRG scheme having a linear (nearest-neighbor) structure, which is an accurate approximation of the much more time-consuming full SDRG scheme for not too small decay exponents $\alpha$.  

By a simple reasoning based on the SDRG method we pointed out that, for large $\alpha$, the true asymptotic behavior is masked by the short-range fixed point at small scales, and derived the dependence of the corresponding crossover length scale on $\alpha$. 
The SDRG scheme also suggests that the effective central charge, as opposed to the nearest-neighbor model, is non-universal, i.e. it depends on the concrete form of the distribution of hopping amplitudes. 

The numerical results showed an overall logarithmic divergence of the entanglement entropy in the extensive regime $\alpha>1$.  
The effective central charge is found to decrease monotonically with $\alpha$; 
 for small enough $\alpha$ it exceeds the effective central charge of the nearest-neighbor model ($\ln 2$), but in the large $\alpha$ limit it seems to saturate to a value which lies below $\ln 2$. 
For $1<\alpha\le 2$, the relationship between $c_{\rm eff}$ and $b$ characteristic for a RSS is found to be violated, while for $\alpha>2$, it is possibly valid but a clear conclusion can not be drawn due to the disturbance of the short-range fixed point which makes strong corrections at the available size scales. 

The SDRG method provides a systematically lower value for $c_{\rm eff}$ than the exact diagonalization, but it becomes more and more accurate with increasing $\alpha$. The SDRG data obtained by the minimal scheme show a dependence of $c_{\rm eff}$ on the distribution of transition amplitudes, which is relatively weak compared to the variation with $\alpha$. This weak non-universality is comparable the precision of ED method.   

Throughout this paper, a variant of the free-fermion model was considered in which hoppings were allowed only from one sublattice to the other, and the question arises how the results obtained here are affected by relaxing this constraint. 
 Applying the SDRG method to general hopping models without a sublattice symmetry a new feature appears, the generation (and decimation) of on-site terms \cite{melin,motrunich}.  
Nevertheless, if we neglect the perturbative corrections involving hopping amplitudes between non-neighboring active sites, then the production of on-site terms is also dropped and we arrive at the same minimal SDRG scheme as was obtained for the bipartite model. From this we conclude that the dominant process in the full SDRG scheme is singlet formation and frozen sites (obtained by on-site decimations) are relatively rare. Indeed, in accordance with this, a vanishing fraction of frozen sites in the tight-binding model on non-bipartite fractal lattices was reported in Ref. \cite{melin}. 
Based on this, we expect the effective central charge of the non-bipartite model to be close to that of the bipartite one, at least for not too small $\alpha$. 

Finally, we mention that through the approximation leading to the minimal SDRG scheme, there is a connection between the hopping model studied in this paper and certain long-range random XX spin models. Keeping exclusively the term involving couplings between neighboring active spins in the perturbative renormalization rule of the XX model (see e.g. Ref. \cite{mohdeb}), we arrive at the form of a renormalized coupling $\tilde J_{14}=w_{14}l_{14}^{-\alpha}+\frac{J_{12}J_{34}}{J_{23}}$, which differs from Eq. (\ref{rule2a}) only by the sign of the second term. For purely antiferromagnetic couplings ($w_{ij}>0$), this is a relevant difference compared to the free-fermion model, since renormalized couplings remain always positive and the two terms on the right-hand side add up always with the same sign. 
If, however, we have an XX spin glass with both ferromagnetic and antiferromagnetic couplings such that the probability density of $w$ is an even function, $f(w)=f(-w)$, then it is easy to see that the difference in the sign of the second term is statistically irrelevant. Thus, such a special long-range XX spin glass has, within the minimal SDRG approximation, the same properties including the effective central charge as the corresponding bipartite hopping model studied in this paper.

\begin{acknowledgments}
The author thanks F. Igl\'oi and G. Ro\'osz for useful discussions. This work was supported by the National Research, Development and Innovation Office NKFIH under Grant No. K128989. 
\end{acknowledgments}

\end{document}